# On Necessary and Sufficient Number of Cops in the Game of Cops and Robber in Multidimensional Grids

Sayan Bhattacharya[1], Goutam Paul[2], Swagato Sanyal[3]

[1] Department of Computer Science,
Duke University, Durham, NC 27708, U.S.A.
`bsayan@cs.duke.edu`
[2] Department of Computer Science and Engineering,
Jadavpur University, Kolkata 700 032, India
`goutam_paul@cse.jdvu.ac.in`
[3] Department of Computer Science and Engineering,
Indian Institute of Technology Kanpur, UP 208016, India
`swagatos@cse.iitk.ac.in`

**Abstract.** We theoretically analyze the Cops and Robber Game for the first time in a multidimensional grid. It is shown that for an $n$-dimensional grid, at least $n$ cops are necessary to ensure capture of the robber. We also present a set of cop strategies for which $n$ cops are provably sufficient to catch the robber. Further, for two-dimensional grid, we provide an efficient cop strategy for which the robber is caught even by a single cop under certain conditions.

## 1 Introduction

The game of 'cops and robber' is played between a number of cops and a single robber on a predefined graph structure. Each of the cops and the robber start from some initial node and moves from one node to another as the game proceeds. The cops win if they can 'capture' the robber in a finite time, otherwise the robber wins. In the literature, there exist different types of movements and various notions of capture (see Section 1.1 for details).

Several attempts have been made to analyze different variants of the cops and robber game in the last two decades. However, there still remain many open problems in this domain, leading to continual research in the topic till date.

### 1.1 Background

Formal investigation into the problem of cops and robber and its variants dates back to early eighties. The works [1,2,3,4,11] consider discrete movements of the cops and the robber in alternate steps, the cops choosing their initial positions first. The robber is assumed to be captured if her position coincides with that of any cop. In [1], an algorithm to determine whether a given graph is *cop-win* is presented (a graph is *cop-win* if a single cop is sufficient to get hold of the robber). In [2], the notion of *cop-number* of a graph (the minimum number of cops needed to ensure that the robber is caught under all possible circumstances) is introduced and a detailed analysis is performed for planar graphs. Later, the works [3,4,11] explored the cop-numbers for different graphs and discovered some interesting bounds.

The game of cops and robber can also be generalized to directed graphs. Here the robber moves with an infinite speed, although he is not permitted to run through a cop. The cops move in helicopters from one node to another node directly, even if the two nodes are not adjacent. Optionally the robber may be considered to be invisible by the cops, but not vice versa. In [5,6], variations of this scheme have been investigated, the main objective being to determine how may cops are necessary to capture the robber.

Another version of the game deals with a two-dimensional grid [7,8,9]. The robber selects her initial position after the cops have chosen theirs and then they keep on moving continuously through the edges of the grid. The robber has complete information about the positions and strategies of all the cops. However, the *visibility* power of each cop is confined to the nodes and edges in her current column (row). The cops win if at some point of time some cop can 'see' the robber. This form of the game has applications in the motion planning of multiple robots [7]. If the robber moves at least as fast as the cops, then according to [7], two cops are necessary and sufficient to ensure a win for the cops. However, only one cop can always catch the robber if she moves fast enough. Subsequent works [8,9] improved the bound on the minimum speed required by a single cop to ensure robber's capture.

The work [10] also considers a two-dimensional grid model, where the cops and the robber choose their initial positions randomly and then move alternately in discrete steps. In addition, the paper discusses applications of this model in the domain of multi-agent systems.

## 1.2 Contribution

To our knowledge, this is the first work on the game of cops and robber in a general $n$-dimensional grid paradigm. Existing relevant works [7,8,9,10] focus only on two-dimensional grid which is a special case with $n = 2$. As an example, one could imagine that the cops are chasing a robber inside a multistoried apartment complex, and model that with a three-dimensional grid and apply our results with $n = 3$. This may find applications in three-dimensional motion planning of robots.

The works [7,8,9] consider that the cops and the robber move simultaneously in a continuous manner. The focus is mainly on the speed requirements and the notion of capture is defined in terms of visibility. On the other hand, we follow the same model as that of [10]. In our work, it is assumed that the cops and the robber choose their initial positions randomly and their movements take place in discrete steps. The robber is considered to be caught if his position coincides with any cop.

In [10], four predator agents (i.e., cops) chase a target agent (i.e. robber) in a square grid. Three related convergence metrics are introduced and an algorithm is presented based on one of them. Applying our general result in two dimensions, only two predator agents can successfully capture the target agent. Thus our work may be considered to be a major improvement over [10].

In Section 2, we rigorously formulate the problem and introduce some terminologies that would be used throughout this paper. Section 3 shows that capture of the robber can never be guaranteed with less than $n$ cops. We also propose a set of cop-strategies and prove that $n$ cops operating in accordance with these strategies will always be able to nab the robber. Section 4 presents a strategy for a single cop in two dimensions. This strategy ensures a win for the cop with a probability of 0.5, provided that the initial positions of the cop and the robber are determined uniformly at random.

## 2 Mathematical Formulation

Let there be $m$ *cops* $C_0, C_1, \ldots, C_{m-1}$ chasing a *robber* $R$. The term *agent* represents either a cop or the robber. Each agent occupies some node of a given undirected graph $G$. A node may contain more than one agent. However, no agent can simultaneously occupy more than one node. Similar to [10], we also assume that the initial positions of the cops and the robber are decided arbitrarily. Whenever an agent moves from one node to an adjacent node, the movement is called a *jump*.

This paper considers the situation when the game is being played on a *n-dimensional grid*. Any node in a $n$-dimensional $d_0 \times d_1 \times \cdots \times d_{n-1}$ grid can be expressed as a $n$ tuple $(u_0, u_1, \ldots, u_{n-1})$ where each $u_i$ is an integer belonging to the closed interval $[0, d_i - 1]$. Two distinct nodes are adjacent if and only if exactly one of their $n$ co-ordinates differ by 1, all other co-ordinates remaining the same. In other words, the nodes $(u_0, u_1, \ldots, u_{n-1})$, $(v_0, v_1, \ldots, v_{n-1})$ are adjacent if and only if $\exists i \in \{0, 1, \ldots, n-1\}$ such that

(a) $\mid u_i - v_i \mid = 1$ and
(b) $\forall j \in \{0, 1, \ldots, n-1\} \setminus \{i\}, u_j = v_j$.

We also assume that $d_i > 1$ for $i = 0, ..., n-1$. Otherwise, a $n$ dimensional grid may degenerate into a lower dimensional grid and some of the results discussed in subsequent sections may no longer be valid.

If an agent occupies some node $(u_0, u_1, \ldots, u_{n-1})$, then $u_j$ would be referred to as the *co-ordinate* $j$ of the current position of that agent, $0 \leq j \leq n-1$. Let $R_j^{(t)}$ denote the value of co-ordinate $j$ of the robber after she completes $t$ jumps and $C_{i,j}^{(t)}$ denote the value of co-ordinate $j$ of cop $C_i$ after her $t^{th}$ jump, $0 \leq j \leq n-1$, $0 \leq i \leq m-1$, $t \geq 0$. The vector $R^{(t)} = (R_0^{(t)}, R_1^{(t)}, \ldots, R_{n-1}^{(t)})$ and the vector $C_i^{(t)} = (C_{i,0}^{(t)}, C_{i,1}^{(t)}, \ldots, C_{i,n-1}^{(t)})$ denote the nodes occupied by the robber $R$ and the cop $C_i$ after their $t$ jumps respectively. Thus, $R^{(0)}$ and $C_i^{(0)}$ denote their initial positions. Whenever the number of jumps is not important, for the sake of simplicity we omit the superscript $t$ and use the notations $R$, $R_j$, $C_i$ and $C_{i,j}$ instead of $R^{(t)}$, $R_j^{(t)}$, $C_i^{(t)}$ and $C_{i,j}^{(t)}$ respectively. It will be clear from the context whether the notations $R$ and $C_i$ denote a particular agent or his/her position. The game continues in the following steps.

1. $t = 0$.
2. The robber jumps to $R^{(t+1)}$.
3. Each cop jumps simultaneously, the new node occupied by cop $C_i$ being $C_i^{(t+1)}$.
4. $t = t + 1$. Go to Step 2.

**Definition 1.** *A* configuration *is defined as the $(m+1)$-tuple $(C_0, C_1, \ldots, C_{m-1}, R)$ and it specifies the position of each cop and the robber at an instant.*

**Definition 2.** *A configuration is* terminating *if some cop occupies the same node as that of the robber.*

**Definition 3.** *A* strategy *for an agent is an algorithm that takes the current configuration as input and returns a node to which the agent will take the next jump.*

While taking a jump, each agent selects an adjacent node by applying his/her own strategy. Our basic objective is to develop strategies for the cops so that eventually some terminating configuration is attained.

**Definition 4.** *A set of strategies for the cops is* winning *if and only if for all possible initial configurations and robber strategies, some terminating configuration is achieved after finite number of jumps.*

## 3 Analysis of the Game in $n$-dimensional Grid

In this section, we formally analyze the minimum number of cops required to ensure capture of the robber in an $n$-dimensional grid. This number is independent of any cop or robber strategy. We also investigate a relevant question: how to construct a set of cop strategies that requires exactly this minimum number of cops and therefore is optimal.

### 3.1 Necessary Number of Cops

We are going to prove that at least $n$ cops are necessary if one wants to guarantee the capture of the robber.

**Definition 5.** $D_{i,j}^{0(t)} = \mid C_{i,j}^{(t)} - R_j^{(t)} \mid$, *and* $D_{i,j}^{1(t)} = \mid C_{i,j}^{(t)} - R_j^{(t+1)} \mid$.
*Moreover,* $D_i^{0(t)} = \sum\limits_{j=0}^{n-1} D_{i,j}^{0(t)}$, *and* $D_i^{1(t)} = \sum\limits_{j=0}^{n-1} D_{i,j}^{1(t)}$.

Note that $D_i^{0(t)}$ is the Manhattan distance between cop $C_i$ and the robber after each of them has taken $t$ jumps. Similarly $D_i^{1(t)}$ is the Manhattan distance between cop $C_i$ and the robber after the robber has taken $(t+1)$ jumps and the cop has taken $t$ jumps. During each jump, an agent (be it the robber or some cop) changes exactly one of its co-ordinates by exactly one unit. We thus have the following result:

**Proposition 1.** *For each cop $C_i$ and $t \geq 0$,*
  (a) $D_i^{1(t)} = D_i^{0(t)} \pm 1$.
  (b) $D_i^{0(t+1)} = D_i^{1(t)} \pm 1$.

**Lemma 1.** *For each cop $C_i$ and $0 \leq t_1 < t_2 < \infty$; $D_i^{0(t_1)}$ and $D_i^{0(t_2)}$ are of same parity.*

*Proof.* From Proposition 1(a), $D_i^{0(t_1)}$ and $D_i^{1(t_1)}$ are of opposite parity. From Proposition 1(b), $D_i^{1(t_1)}$ and $D_i^{0(t_1+1)}$ are of opposite parity. So $D_i^{0(t_1)}$ and $D_i^{0(t_1+1)}$ are of same parity. Similarly we may prove that $D_i^{0(t_1+1)}$ and $D_i^{0(t_1+2)}$ are of same parity. Continuing in this manner, we reach the desired result. □

Next, we are going to show (Theorem 1) that in an $n$-dimensional grid, there does not exist any winning set of strategies (Definition 4) for less than $n$ cops.

**Theorem 1.** *In a $n$-dimensional grid, if the initial configuration is such that $D_i^{0(0)}$ is of odd parity for each cop $C_i$, then there exists some robber strategy for which the robber can never be caught with less than $n$ cops.*

*Proof.* Lemma 1 implies that $\forall t > 0$ and $\forall i \in \{0, 1, \ldots, m-1\}$, $D_i^{0(t)}$ is of odd parity, $m$ being the number of cops. The configuration attained after the robber and all the cops have each taken $t$ jumps can never be terminating; otherwise, there would be some cop $C_j$ occupying the same node with robber, implying that $D_j^{0(t)} = 0$, an integer with even parity.

Since $R^{(t)}$ has at least $n$ adjacent nodes and there are less than $n$ cops, the robber may easily jump to a node which is not occupied by any cop. Accordingly the configuration attained here (after the robber has taken $t+1$ jumps and each cop has taken $t$ jumps) is also not terminating. □

**Corollary 1.** *In a $n$-dimensional grid, $n$ cops are necessary to guarantee capture of the robber.*

### 3.2  Winning Set of Strategies and Sufficient Number of Cops

Corollary 1 poses a natural question for $n$-dimensional grid: does there exist a wining set of strategies for exactly $n$ cops? Before addressing this issue, we present some general results that hold for all cop strategies. These results would be required for subsequent analysis in this section.

**Definition 6.** $J_R^{(t)} \doteq$ *the co-ordinate along which robber makes a move in his $t^{th}$ jump.* $J_{C_i}^{(t)} \doteq$ *the co-ordinate along which cop $C_i$ makes a move in her $t^{th}$ jump.*

We use the notation $a \bullet b$ to denote $(a+b) \bmod n$, where $a$ and $b$ are integers.

**Definition 7.** *For all $t \geq 0$, the $(t+1)^{th}$ jump of the robber is* favorable *to cop $C_i$ if and only if $J_R^{(t+1)} \notin \{i, i \bullet 1, \ldots, i \bullet (j_1 - 1)\}$. Here $j_1$ is the smallest integer in the set $\{0, 1, \ldots, n-1\}$ which satisfies the inequality $C_{i,i \bullet j_1}^{(t)} \neq R_{i \bullet j_1}^{(t)}$.*

Observe that if $C_{i,i}^{(t)} \neq R_i^{(t)}$, then $j_1 = 0$ and $\{i, i \bullet 1, \ldots, i \bullet (j_1 - 1)\}$ reduces to the null set. In such cases the next jump of the robber will always be favorable to cop $C_i$.

**Lemma 2.** *For $p = 0, 1, \ldots, n-1$, if the robber moves along co-ordinate $p$ in his $(t+1)^{th}$ jump, then that jump is favorable to cop $C_{p \bullet 1}$.*

*Proof.* We can safely assume that $C_{p \bullet 1}^{(t)} \neq R^{(t)}$ (Otherwise the robber has already been captured). Let $j_1$ be the minimum integer in the set $\{0, 1, \ldots, n-1\}$ such that $C_{p \bullet 1, (p \bullet 1) \bullet j_1}^{(t)} \neq R_{(p \bullet 1) \bullet j_1}^{(t)}$. Now, $J_R^{(t+1)} = p = (p \bullet 1) \bullet (n-1) \notin \{(p \bullet 1), (p \bullet 1) \bullet 1, \ldots, (p \bullet 1) \bullet (j_1 - 1)\}$, since $j_1 - 1 \leq n - 2$. By Definition 7, the $(t+1)^{th}$ jump of the robber is favorable to cop $C_{p \bullet 1}$. □

Algorithm 1 shows a set of cop strategies that would later be proved to be winning. The strategy for each cop $C_i$ is denoted by $S_i$. A major advantage of this set of strategies is that a cop need not know the positions of other cops. $S_i$ determines $C_i^{(t+1)}$ based on $C_i^{(t)}$ and $R^{(t+1)}$ only. The purpose of the loop in Step 2 is to find an integer $j \in \{0, 1, \ldots, n-1\}$ such that $C_{i,i}^{(t)} = R_i^{(t+1)}, C_{i,i \bullet 1}^{(t)} = R_{i \bullet 1}^{(t+1)}, \ldots, C_{i,i \bullet (j-1)}^{(t)} = R_{i \bullet (j-1)}^{(t+1)}$ and $C_{i,i \bullet j}^{(t)} \neq R_{i \bullet j}^{(t+1)}$. Either such an integer $j$ exists, or $C_i^{(t)} = R^{(t+1)}$. In Step 3, $j = n$ if and only if $C_i^{(t)} = R^{(t+1)}$, indicating that a terminating configuration has already been achieved. Step 4 determines the node to which cop $C_i$ is going to jump in case the present configuration is not a terminating one.

| **Algorithm 1: Strategy $S_i$ for cop $C_i$** |
|---|
| 1.      $j \leftarrow 0$; |
| 2.      While $(j \neq n)$ |
| 2.1.          If $R_{i \bullet j}^{(t+1)} \neq C_{i,i \bullet j}^{(t)}$ then go to Step 3; |
| 2.2.          $j \leftarrow j + 1$; |
| 3.      If $j = n$ then terminate the game; |
| 4.      Else |
| 4.1.          $C_i^{(t+1)} \leftarrow C_i^{(t)}$; |
| 4.2.          If $R_{i \bullet j}^{(t+1)} < C_{i,i \bullet j}^{(t)}$ then $C_{i,i \bullet j}^{(t+1)} \leftarrow C_{i,i \bullet j}^{(t+1)} - 1$; |
| 4.3.          Else $C_{i,i \bullet j}^{(t+1)} \leftarrow C_{i,i \bullet j}^{(t+1)} + 1$; |

**Lemma 3.** *In a n-dimensional $d_0 \times d_1 \times \cdots \times d_{n-1}$ grid, if the robber has taken $\sum_{j=0}^{n-1} d_j$ jumps favorable to cop $C_i$ and the cop follows strategy $S_i$, then a configuration has already been attained where the robber and cop $C_i$ occupied the same node.*

*Proof.* We are going to show that if the robber has taken $(d_i + d_{i\bullet 1} + \cdots + d_{i\bullet k})$ jumps favorable to cop $C_i$, then earlier a configuration was reached where $R_i = C_{i,i}, R_{i\bullet 1} = C_{i,i\bullet 1}, \ldots, R_{i\bullet k} = C_{i,i\bullet k}$. We use induction on $k$.

*Base step* ($k = 0$): Without any loss of generality, let $R_i^{(0)} > C_{i,i}^{(0)}$. Until $R_i = C_{i,i}$, every jump of the robber is favorable to cop $C_i$. At each jump, value of $(R_i - C_{i,i})$ changes by at most 1. Initially $(R_i - C_{i,i})$ is positive. It cannot become negative without touching 0 at some stage. And it cannot remain positive for more than $d_i$ jumps of the robber. Otherwise $C_{i,j}$ is constantly incremented by 1 for more than $d_i$ times, a contradiction.

*Induction step*: Let the statement be true for $k = l$. We start from the configuration where $R_i = C_{i,i}, R_{i\bullet 1} = C_{i,i\bullet 1}, \ldots, R_{i\bullet l} = C_{i,i\bullet l}$. Let $R_{i\bullet(l+1)} > C_{i,i\bullet(l+1)}$. If a jump of the robber is not favorable to cop $C_i$, then the subsequent jump of the cop is used to maintain the above mentioned equalities. Otherwise $C_{i,i\bullet(l+1)}$ is adjusted so as to get it closer to $R_{i,i\bullet(l+1)}$. Similarly as in the *Base step*, it may be shown that a configuration, where $R_i = C_{i,i}, R_{i\bullet 1} = C_{i,i\bullet 1}, \ldots, R_{i\bullet l} = C_{i,i\bullet l}$ and $R_{i\bullet(l+1)} = C_{i,i\bullet(l+1)}$, will be attained within $(d_i + d_{i\bullet 1} + \cdots + d_{i\bullet l} + d_{i\bullet(l+1)})$ favorable robber-jumps.

Putting $k = n - 1$, the result follows. □

**Theorem 2.** *In an n-dimensional grid, n cops are sufficient to ensure capture of the robber.*

*Proof.* Suppose each cop $C_i$ follows strategy $S_i$. By Lemma 2, each jump of the robber is favorable to some cop. By Pigeonhole principle, if the robber takes $n \sum_{i=0}^{n-1} d_i$ jumps, then at least $\sum_{i=0}^{n-1} d_i$ of them would be favorable to some specific cop. By Lemma 3, this implies that a terminating configuration has been reached. Thus, the set of cop strategies $\{S_i \mid 0 \leq i < n\}$ guarantees capture of the robber. □

Since, for each of $n \times \sum_{i=0}^{n-1} d_i$ jumps of the robber, the $n$ cops take simultaneous jumps in $O(n)$ time, the worst case run-time of the set of cop strategies of Algorithm 1 is $O(n^2 \sum_{i=0}^{n-1} d_i)$.

The set of strategies outlined in Algorithm 1 is optimal in the sense that they guarantee the attainment of a terminating configuration using minimum number of cops.

### 3.3 Some Experimental Results for Two Dimensions

The robber cannot ensure evasion in two dimensions if two cops are chasing her. But she may want to delay her capture. We empirically observe how many jumps are taken by the robber before she is caught. We consider three different robber strategies (assuming that exactly two cops are present) described below. The cops move in accordance with the winning set of strategies presented in this section.

**Robber Strategy 1**: For each adjacent position $(x, y)$, she evaluates the expression $\{(x - C_{0,0})^2 + (y - C_{0,1})^2\} + \{(x - C_{1,0})^2 + (y - C_{1,1})^2\}$ and moves to that adjacent position for which the expression is maximum.

**Robber Strategy 2**: For each adjacent position $(x, y)$, she evaluates the expression $\{|x - C_{0,0}| + |y - C_{0,1}|\} + \{|x - C_{1,0}| + |y - C_{1,1}|\}$ and moves to that adjacent position for which the expression is maximum.

**Robber Strategy 3**: For each adjacent position $(x, y)$, she evaluates the expression $\sqrt{(x - C_{0,0})^2 + (y - C_{0,1})^2} + \sqrt{(x - C_{1,0})^2 + (y - C_{1,1})^2}$ and moves to that adjacent position for which the expression is maximum.

Table 1 shows the average number of jumps for the three different robber strategies when the game is repeated 1,000,000 times, each time with a random initial configuration.

| Grid Size | Average No. of Jumps by Robber | | |
|---|---|---|---|
| | Strategy 1 | Strategy 2 | Strategy 3 |
| 10 x 10 | 8 | 11 | 13 |
| 15 x 15 | 14 | 17 | 23 |
| 20 x 20 | 19 | 24 | 32 |
| 25 x 25 | 24 | 31 | 41 |
| 30 x 30 | 30 | 37 | 50 |
| 35 x 35 | 35 | 44 | 59 |
| 40 x 40 | 40 | 51 | 69 |
| 45 x 45 | 46 | 57 | 78 |
| 50 x 50 | 51 | 64 | 87 |

**Table 1.** Performance comparison of three different robber strategies.

As the table shows, strategy 3 seems to be most effective for the robber.

## 4 Additional Theoretical Results Specific to Two Dimensions

Consider that a single cop is chasing the robber in two dimensions. According to Theorem 1, the robber can always evade capture for certain *bad* initial configurations (where $D_0^{0(0)}$ is odd). If the starting positions of the cop and the robber are chosen uniformly at random, then the probability that a bad initial configuration will be encountered is 0.5. In all other situations, the cop-strategy $S$ presented in Algorithm 2 guarantees capture of the robber.

**Algorithm 2: Strategy $S$ in two dimensions**

1. $C_0^{(t+1)} \leftarrow C_0^{(t)}$;
2. If $\mid C_{0,0}^{(t)} - R_0^{(t+1)} \mid > \mid C_{0,1}^{(t)} - R_1^{(t+1)} \mid$ then
2.1     If $C_{0,0}^{(t)} > R_0^{(t+1)}$ then $C_{0,0}^{(t+1)} \leftarrow C_{0,0}^{(t)} - 1$;
2.2     Else $C_{0,0}^{(t+1)} \leftarrow C_{0,0}^{(t)} + 1$;
3. Else if $\mid C_{0,0}^{(t)} - R_0^{(t+1)} \mid < \mid C_{0,1}^{(t)} - R_1^{(t+1)} \mid$ then
3.1     If $C_{0,1}^{(t)} > R_1^{(t+1)}$ then $C_{0,1}^{(t+1)} \leftarrow C_{0,1}^{(t)} - 1$;
3.2     Else $C_{0,1}^{(t+1)} \leftarrow C_{0,1}^{(t)} + 1$;
4. Else jump to any adjacent node;

Definition 5 takes the following form in a two-dimensional grid: $D_0^{0(t)} = D_{0,0}^{0(t)} + D_{0,1}^{0(t)} = \mid C_{0,0}^{(t)} - R_0^{(t)} \mid + \mid C_{0,1}^{(t)} - R_1^{(t)} \mid$, $D_0^{1(t)} = D_{0,0}^{1(t)} + D_{0,1}^{1(t)} = \mid C_{0,0}^{(t)} - R_0^{(t+1)} \mid + \mid C_{0,1}^{(t)} - R_1^{(t+1)} \mid$.

**Proposition 2.** *If $\mid C_{0,0} - R_0 \mid$ becomes equal to $\mid C_{0,1} - R_1 \mid$ at a stage when both the cop and the robber have taken same number of jumps, then from that point onwards the sign of $C_{0,0} - R_0$ (as well as that of $C_{0,1} - R_1$) does not change, provided the cop moves in accordance with strategy $S$.*

*Proof.* Let for some $t$, $\mid C_{0,0}^{(t)} - R_0^{(t)} \mid = \mid C_{0,1}^{(t)} - R_1^{(t)} \mid$. Without any loss of generality, let $C_{0,0}^{(t)} < R_0^{(t)}$ and $C_{0,1}^{(t)} > R_1^{(t)}$. The robber may now either increment (decrement) $R_0$, or she may increment (decrement) $R_1$. Irrespective of the choice she makes, the cop following strategy $S$ will move in such a fashion so as to maintain the equality $\mid C_{0,0}^{(t+1)} - R_0^{(t+1)} \mid = \mid C_{0,1}^{(t+1)} - R_1^{(t+1)} \mid$. Further, we shall have $C_{0,0}^{(t+1)} \leq R_0^{(t+1)}$ and $C_{0,1}^{(t+1)} \geq R_1^{(t+1)}$. In other words, either $C_{0,0}^{(t+1)} = R_0^{(t+1)}$, $C_{0,1}^{(t+1)} = R_1^{(t+1)}$ and the game terminates; or we have $C_{0,0}^{(t+1)} < R_0^{(t+1)}$ and $C_{0,1}^{(t+1)} > R_1^{(t+1)}$. The same line of reasoning may be repeated arbitrary number of times. □

**Lemma 4.** *Consider an initial configuration such that $D_0^{0(0)}$ is of even parity. After the cop, who follows strategy $S$, moves along co-ordinate $i$ for the first time, the sign of the expression $C_{0,i} - R_i$ is never going to change, for $i = 0, 1$.*

*Proof.* We validate the above statement only for co-ordinate 0. The other possible case for co-ordinate 1 can be proved in a similar way. Let the cop move along co-ordinate 0 for the first time in her $(t_1 + 1)^{th}$ jump. Since $D_0^{0(0)}$ is even, $\mid C_{0,0}^{(t_1)} - R_0^{(t_1+1)} \mid - \mid C_{0,1}^{(t_1)} - R_1^{(t_1+1)} \mid$ must be odd and hence nonzero. According to strategy $S$, $\mid C_{0,0}^{(t_1)} - R_0^{(t_1+1)} \mid - \mid C_{0,1}^{(t_1)} - R_1^{(t_1+1)} \mid > 0$. Without any loss of generality, we assume that $C_{0,0}^{(t_1)} < R_0^{(t_1+1)}$ and the cop increments $C_{0,0}$ in her $(t_1 + 1)^{th}$ jump.

If possible, let the sign of $C_{0,0} - R_0$ become positive at some point of time after the cop has taken her $(t_1+1)^{th}$ jump. But prior to that, $C_{0,0} - R_0$ must touch the value 0; for $C_{0,0} - R_0$ changes by at most 1 during each step. When $C_{0,0} - R_0 = 0$, $\mid C_{0,0} - R_0 \mid - \mid C_{0,1} - R_1 \mid \leq 0$. The value of $\mid C_{0,0} - R_0 \mid - \mid C_{0,1} - R_1 \mid$ also changes by at most 1 during each jump of the cop or robber. Consequently the game must have gone through a stage where $\mid C_{0,0} - R_0 \mid - \mid C_{0,1} - R_1 \mid = 0$ and $C_{0,0} \leq R_0$. Moreover, this particular stage must have been attained after the $(t_1+1)^{th}$ jump by the cop and prior to the moment when $C_{0,0} - R_0$ becomes positive for the first time. Since $D_0^{0(0)}$ was even, both the cop and the robber must have taken same number (say $t_2$) of jumps before reaching the above mentioned stage. If $C_{0,0}^{(t_2)} = R_0^{(t_2)}$, then the game terminates immediately, ruling out the option for $C_{0,0} - R_0$ to become positive. Else if $C_{0,0}^{(t_2)} < R_0^{(t_2)}$, we apply Proposition 2 to show that $C_{0,0}^{(t_2)} - R_0^{(t_2)}$ will never be positive as the game proceeds. This leads to a contradiction. $\square$

**Theorem 3.** *If the initial configuration is such that $D_0^{0(0)}$ is even, then the strategy presented in Algorithm 2 ensures a win for the cop.*

*Proof.* If possible, let $D_0^{0(0)}$ be even and still the cop fails to nab the robber. Lemma 4 implies that the cop will never backtrack along any of its co-ordinates. Moreover, the cop has to take an infinite number of jumps. Since we only consider finite grids, this leads to a contradiction. $\square$

**Theorem 4.** *The cop-strategy in Algorithm 2 succeeds in capturing the robber on average half the times the game is repeated, given that the initial positions of the cop and the robber are decided uniformly at random.*

*Proof.* By definition, $D_0^{0(0)} = D_{0,0}^{0(0)} + D_{0,1}^{0(0)} = \mid C_{0,0}^{(0)} - R_0^{(0)} \mid + \mid C_{0,1}^{(0)} - R_1^{(0)} \mid$. $C_{0,0}^{(0)}$, $R_0^{(0)}$, $C_{0,1}^{(0)}$, and $R_1^{(0)}$ are each chosen uniformly at random. Hence each of these is expected to be odd (or even) half of the times, and so will be each of $\mid C_{0,0}^{(0)} - R_0^{(0)} \mid$ and $\mid C_{0,1}^{(0)} - R_1^{(0)} \mid$, and their sum. Now the result follows immediately from Theorem 3. $\square$

If $D_0^{0(0)}$ is even, then the cop always moves in a fixed direction along co-ordinate 0 (as well as along co-ordinate 1). The robber will be caught within $O(d_0 + d_1)$ jumps of the cop (recall that the game is being played in a $d_0 \times d_1$ grid). A cop following Algorithm 2 can decide in constant time where to jump. Hence the time complexity of strategy S is also $O(d_0 + d_1)$. Here we exclude all initial configurations with an odd value of $D_0^{0(0)}$; as the robber can perpetually evade capture in such cases.

Now consider the situation where initially the cop and the robber are situated at diagonally opposite corners of the grid and the robber's strategy dictates him to stay as close as possible to his initial position. Obviously the cop will have to take at least $O(d_0 + d_1)$ jumps to catch the robber. We thus have the following result.

**Theorem 5.** *Unless the initial configuration is such that the robber has the privilege to evade capture indefinitely, the cop-strategy in Algorithm 2 ensures a win for the cop in asymptotically optimal time.*

## 5 Conclusion

We analyze the Cops and Robber Game in a $n$-dimensional grid structure and show that $n$ cops are both necessary and sufficient to capture the robber. We present a set of cop strategies which satisfies this sufficiency condition. Moreover, in two-dimensional grid, we show that even a single cop can catch the robber under certain cases. In our future work, we plan to investigate whether such strategies exist in general in $n$ dimensions, i.e., strategies that would guarantee capture of the robber in some special cases with less than $n$ cops.

**Acknowledgment:** The authors like to thank Mr. Somitra Sanadhya, Indian Statistical Institute, Kolkata for introducing them to the game of Cops and Robber for the first time.